\documentclass[prb,amsmath,superscriptaddress,citeautoscript,twocolumn,
showpacs,floatfix]{revtex4}
\usepackage{graphicx}

\usepackage{amssymb}
\usepackage{color}
\usepackage{epsfig}
\def\cals{{\mbox{\boldmath{${\cal S}$}}}}

\def\sig{{\mbox{\boldmath{$\sigma$}}}}

\begin{document}

\def\sigmav{{\mbox{\boldmath{$\sigma$}}}}

\title{Spin-polarized electric currents in quantum  transport \\ through tubular two-dimensional electron gases}

\author{O. Entin-Wohlman}
\altaffiliation{Also at Tel Aviv University, Tel Aviv 69978,
Israel}

\affiliation{Department of Physics and the Ilse Katz Center for
Meso- and Nano-Scale Science and Technology, Ben Gurion
University, Beer Sheva 84105, Israel}

\affiliation{Albert Einstein Minerva Center for Theoretical
Physics, Weizmann Institute of Science, Rehovot 76100, Israel}


\author{A. Aharony}

\altaffiliation{Also at Tel Aviv University, Tel Aviv 69978,
Israel}

\affiliation{Department of Physics and the Ilse Katz Center for
Meso- and Nano-Scale Science and Technology, Ben Gurion University, Beer
Sheva 84105, Israel}

\author{Y. Tokura}

\affiliation{NTT Basic Research Laboratories, NTT Corporation,
Atsugi-shi, Kanagawa 243-0198, Japan}

\author{Y. Avishai}

\affiliation{Department of Physics and the Ilse Katz Center for
Meso- and Nano-Scale Science and Technology, Ben Gurion
University, Beer Sheva 84105, Israel}

\date{\today}
\begin{abstract}

Scattering theory is employed to derive
a Landauer-type formula for the spin and the charge  currents, through a finite region where spin-orbit interactions are effective. It
is shown that the transmission matrix yields the spatial direction
and the magnitude of the spin polarization. This formula is used to study the currents through a tubular two-dimensional electron gas. In this cylindrical geometry, which may be realized in experiment,  the transverse  conduction
channels are not mixed (provided that the spin-orbit coupling is uniform). It is then found that  for modest boundary scattering, each
step in the quantized conductance is split into two, and the new
steps have a non-zero spin conductance, with the spin polarization
perpendicular to the direction of the current.

\end{abstract}


\maketitle

\section{Introduction}

The conductance of a confined system connected to    electronic reservoirs
(held at slightly different chemical potentials) can be expressed
by its scattering properties, notably the transmission,
leading to the celebrated Landauer formula \cite{landauer,LANDAUER,book}.
When the motion of the electrons is
ballistic and the transmission is perfect, the two-terminal conductance  is
given by
\begin{align}
{\cal G}=\frac{2e^{2}}{h}N_{\perp}\ ,\label{BALLI}
\end{align}
where $N_{\perp}$ is the number of conducting channels  not
including spin.   As a function of the energy of the injected
electrons, additional channels  open successively and the
conductance follows a staircase structure, as was indeed observed
in experiment \cite{wharam}. When the motion is not ballistic, Eq.
(\ref{BALLI}) is modified, with the number of channels being
replaced by the sum over the transmissions among channels
\cite{book,buttiker}.

Here we extend this picture to  include spin-orbit interactions, and derive both  charge  and  spin conductances in terms of the  scattering matrix of the mesoscopic system. Our approach is
based on scattering theory in which the entire
system is represented by its scattering matrix. \cite{buttiker,levinson} We avoid
the ambiguities associated
with the definition of  spin currents in systems with spin-orbit interactions
\cite{niu} by considering scattering through a finite region where these interactions are effective, and computing the currents
far away from that region,  where these interactions are absent. \cite{NUD}
In this respect our approach
differs from those presented in Refs. ~\onlinecite{MOROZ} and ~\onlinecite{PRIVMAN}, which have deduced the conductance from  the energy spectrum, or  of  Ref.~
\onlinecite{silvestrov}, which  employed semiclassical
arguments to study
 spin polarized currents in smooth barriers.

When the  electrons are restricted to move in a plane, an asymmetry
in the confining potential leads to the appearance of the Rashba
spin-orbit interaction. \cite{rashba} Another source for the
spin-orbit coupling is the Dresselhaus mechanism, \cite{DRES} in
particular the cubic one, which might be quite substantial in GaAs
nano-structures. \cite{Bert,CUBIC}  The Rashba spin-orbit interaction  is  in particular interesting,    since its strength  can be controlled in
experiment, by applying  gate voltages. \cite{nitta,SO1,SO2}  From the strictly theoretical point of
view, however,  there is not much difference between the {\em linear}
Dresselhaus interaction and the Rashba term, as they are connected
by a unitary transformation. \cite{sheka}
The  scattering formalism presented in Sec. \ref{LAN}  is independent of the type of the spin-orbit coupling.

Spin-orbit interactions, which couple the momentum of the electron
to its spin, have  attracted much interest due to the possibility
to manipulate the spin  by electric fields. When electrons are injected from spin-polarized electrodes into a spin-orbit coupled region, they might {\em lose} their polarization; this decoherence effect has been studied in Ref. ~\onlinecite{refereeb}. The role of lateral
interfaces formed between regions with different  spin-orbit
couplings \cite{sablikov} in {\em polarizing} the spins, and its analogy
with optics,  has been established  in Refs.~ \onlinecite{khodas}
and ~\onlinecite{BERCIOUX}. However, as direct measurements of the
electron spin polarization or of the electron spin accumulation
are not easily accomplished (see however Ref. ~\onlinecite{halperin1}), it may be helpful to study the effect
of that polarization on the (more easily accessed)  electronic
charge conductance. At the same time, it is desirable to analyze
the spin currents  which arise upon the application  of a small
source-drain bias and explore their possible effect on the charge
transport. Previous studies of the conductance through a region where spin-orbit interactions are active had relied on numerical approximations, \cite{MIRELES,ZHU,Governale,ETO} and consequently some of the conclusions drawn were dictated by a specific choice of parameters.

Here we study coherent transport through a mesoscopic hollow cylinder; this geometry enables an exact calculation of the scattering matrix, and reveals its symmetries. In particular we are able to analyze the effect of these symmetries on the  spin polarization of the transport current.
We present in Sec.  \ref{LAN} our scattering formalism  for spin-dependent transport which follows closely the derivations of the Landauer formula for spin-independent potentials as formulated in Refs. ~\onlinecite{buttiker} and ~\onlinecite{levinson}. We  identify there  the specific combinations of
the elements of the scattering matrix that convey the information on the spin-polarization direction and  its magnitude.   By carrying out this analysis we are able to show that
the transmission matrix of the finite system gives directly the
direction and the amount of the spin polarization. Moreover, under
not too stringent conditions the conductance itself carries
information on the spin polarization. In Sec. \ref{CYLINDER} we study the implications of this generalized Landauer formula for a specific system: a coherent mesoscopic hollow cylinder.
This system, whose experimental realization is found in  the carbon nanotubes, \cite{ANDO}  has recently attracted much interest since the remarkable observation of the effect of the electron motion on the direction of its spin. \cite{SHACHAL} 
Another experimental realization of this geometry may be found in the 
core-shell nanowires. \cite{CORE}
Finally, we summarize our conclusions in Sec. \ref{DISCU}.

\section{Landauer formula for spin-dependent transport}

\label{LAN}

In scattering theory \cite{buttiker,levinson} the electron field operator (a spinor) is expressed
 in terms of the scattering states of energy
$E$,
\begin{align}
\Psi ({\bf r},t)=\int\frac{dE}{2\pi} e^{-iE t}\sum_{a
n \eta}c_{a n \eta}^{}(E )\chi^{}_{a n \eta }({\bf
r};E )\ .\label{FIELD}
\end{align}
(Here and
below $\hbar
=1$.)
In Eq. (\ref{FIELD}), $\chi^{}_{a n\eta}({\bf r};E )$ is a solution
of the Schr\"{o}dinger equation
\begin{align}
(E-{\cal H})\chi_{an\eta}^{}({\bf r};E )=0\ ,
\end{align}
with ${\cal H}$ being the Hamiltonian of the entire system. It
is the scattering state excited by an electron of spin polarization $\eta$ incoming  from  the $n$th channel of terminal $a$. The
operator
$c^{}_{a n \eta}(E )$ $[c^{\dagger}_{a n \eta}(E )]$ destroys (creates) an
electron in such a state.
The thermal average of these operators
(denoted by a bar) is defined by the temperature $T$ and the chemical
potential of the reservoirs. For reservoirs of un-polarized
electrons,
\begin{align}
\overline{ c^{\dagger}_{an\eta}(E )c_{a'n'\eta '}(E
')} =2\pi\delta (E -E ')\delta^{}_{aa'}\delta^{}_{nn'}\delta^{}_{\eta\eta '}f_{a}(E )\ ,\label{CAVE}
\end{align}
where
\begin{align}
f_{a}(E )=(e^{(E -\mu_{a})/k_{B}T}+1)^{-1}
\end{align}
is
the Fermi distribution with the chemical potential $\mu_{a}$ of
reservoir $a$. \cite{COMMENT1}
Equations (\ref{FIELD})
and (\ref{CAVE}) are used to
obtain the thermal averages of the charge and the spin currents.  This is accomplished by writing down  the currents in terms of the field  $\Psi$, Eq. (\ref{FIELD}), and then performing the thermal average over the creation and destruction operators, ${c}$ and $c^{\dagger}$, according to Eq. (\ref{CAVE}). \cite{buttiker,levinson}

To avoid cumbersome notations, we confine ourselves to the case of two terminals, located for concreteness along the $x$ direction. Then, all currents flow along $x$ and in the absence of a magnetic field  are given by
\begin{align}
&I_{x}^{j}({\bf r})=\int\frac{dE}{2\pi}\sum_{an\eta}f_{a}(E )\nonumber\\
&\times\frac{1}{2m}\Bigl (\langle \chi^{}_{an\eta
}({\bf r};E )| \sigma^{}_{j}|(-i\frac{\partial}{\partial
x} ) \chi^{}_{an\eta}({\bf r};E )\rangle+{\rm cc}\Bigr )\
,\label{IRMF}
\end{align}
where
$\sigma^{}_{j}$ is the  $j$th Pauli
matrix. The charge current, $I_{x}$, is given by
Eq. (\ref{IRMF}) upon identifying
$\sigma_{j=0}$ as the unit matrix
(and multiplying by the electronic charge $e$). For
a two-terminal system, the lead index is $a=L$ or $R$, for the left and for the
right reservoirs, respectively.

The currents given by Eq. (\ref{IRMF})  are computed
far away from the scattering region, where they are uniform,
e.g., on the left side.  To this end we represent the scattering states by the incoming and outgoing exciting
waves, e.g., 
\begin{align}
&\chi_{an\eta}(x\ {\rm in}\ L, y,z;E
)=\delta^{}_{a,L}\omega^{\rm in}_{an\eta}({\bf r};E
)\nonumber\\
&+\sum_{n'\eta '}\omega^{\rm out}_{L n'\eta '}({\bf r};E
){\cal S}^{}_{L n'\eta ',a n\eta}(E)\ .\label{YE}
\end{align}
Here, $\omega_{an\eta}^{\rm in, out}$ are the exciting waves,
incoming or outgoing in channel $n$ of lead $a$ and having the
spin polarization $\eta$, normalized to
carry a unit flux, and ${\cal S}$  is the scattering
matrix of our system. The meaning of Eq. (\ref{YE}) is quite transparent: the scattering
state which has been excited by  a wave incoming in lead $a$
consists, when considered in the left lead $L$, of all waves that are
scattered into that lead from lead $a$ (including those that are
reflected), as represented by the second term in Eq. (\ref{YE}), and
the incoming wave in $a$, in case $a$ coincides with $L$.
\cite{levinson}

Next, the expansion (\ref{YE}) is inserted into Eq. (\ref{IRMF}), and
the integration over the cross section is performed.
As a result, $I_{x}^{j}$ becomes spatially-independent,
\begin{align}
&I_{x}^{j}=\int\frac{dE}{2\pi}\sum_{a}f_{a}(E
) {\rm Tr}\Bigl \{ \delta^{}_{a,L }\sigma^{}_{j}
-\sum_{nn'}{\cal M}^{}_{Ln',an}(E)\sigma^{}_{j}
\Bigr \}\nonumber\\
&=\int\frac{dE}{2\pi}\Bigl (f^{}_{L}(E)-f^{}_{R}(E)\Bigr )
{\rm Tr}\sum_{nn'}{\cal M}^{}_{Ln',Rn}(E)\sigma^{}_{j}
\ ,\label{IR1MF}
\end{align}
where  the trace is carried out in spin space. Here
we have introduced the definition
\begin{align}
{\cal M}_{Ln',an}(E)\equiv\cals^{}_{Ln', an}(E)\cals^{\dagger}_{Ln',an}(E)\ ,\label{CALM}
\end{align}
where $\cals_{L n',Rn}$ denotes an entry of the scattering matrix which is a 2$\times$2
matrix in the spin space.
The second equality in Eq. (\ref{IR1MF}) is derived from the unitarity of the scattering matrix. Note that $I^{j}_{x}$ vanishes when there is no bias voltage, $\mu^{}_L=\mu^{}_R$. Although it appears as if pertains for general values of the bias voltage, Eq. (\ref{IR1MF}) is strictly valid only in the linear-response regime, in where $\mu_{L}-\mu_{R}$ approaches zero. A detailed account of the subtleties associated with the Landauer formula can be found in Ref. ~\onlinecite{LANDAUER}.

The general expression Eq. (\ref{IR1MF}) for the charge and the spin currents is our
central result in this section. In the small-bias limit it yields the linear
charge and spin conductances. To express those, it is useful to
note that
the transmission from channel $n $ to channel $n'$ is determined by the matrix ${\cal M}$, Eq. (\ref{CALM}). Since this matrix is obviously hermitian, it can be decomposed into a scalar and a vector, \cite{refereeb}
\begin{align}
{\cal M}^{}_{Ln',Rn}(E)=\frac{1}{2}\Bigl ({\cal T}^{}_{n'n}(E)+{\bf V}^{}_{n'n}(E)\cdot\sig\Bigr )\ ,\label{NEW}
\end{align}
where both
${\cal T}_{n'n}$ and   ${\bf V}_{n'n}$ are real.
The two  real eigenvalues of this matrix are
\begin{align}
\lambda^{\pm}_{n'n}(E)=\frac{1}{2}\Bigl ({\cal T}^{}_{n'n}(E)\pm |{\bf V}^{}_{n'n}(E)|\Bigr )\ .\label{MG}
\end{align}
Their corresponding  eigenvectors are spinors fully polarized along $\pm{\bf V}_{n'n}$.
The total transmission between these two channels is given by
\begin{align}
{\cal T}_{n'n}={\rm Tr}[{\cal M}^{}_{Ln',Rn}]=\lambda^{+}_{n'n}+\lambda^{-}_{n'n}\ , \label{TT}
\end{align}
while the spin transmission  is
\begin{align}
{\rm Tr}[{\cal M}^{}_{Ln',Rn} \sig]={\bf V}_{n'n} \ , \label{VV}
\end{align}
with
\begin{align}
|{\bf V}_{n'n}|=\lambda^{+}_{n'n}-\lambda^{-}_{n'n}\ .\label{VG}
 \end{align}
As is well known, (see e.g., Refs. ~\onlinecite{silvestrov} and ~\onlinecite{khodas})
spin-orbit interactions may turn one of the modes (for each energy) to be evanescent; when this happens, the eigenvalue belonging to that mode, $\lambda^{-}$, is vanishingly small. As a result,  the (perfect) conductance is reduced to $\lambda^{+}\simeq 1$ (in units of $e^2/h$), while the current becomes almost  fully spin-polarized, its respective conductance being
$|{\bf V}_{nn'}|\simeq\lambda^{+}_{nn'}\simeq 1$.
On the other hand, when none of the modes is evanescent, one has $\lambda^{+}_{nn'}\simeq\lambda^{-}_{nn'}\simeq 1$, leading to ${\cal G}=2e^{2}/h$ and zero spin polarization for the perfect conductor.

\section{Charge and spin transport through a coherent  tubular two-dimensional electron gas}
\label{CYLINDER}

\begin{figure}[h ]
\includegraphics[width=8cm]{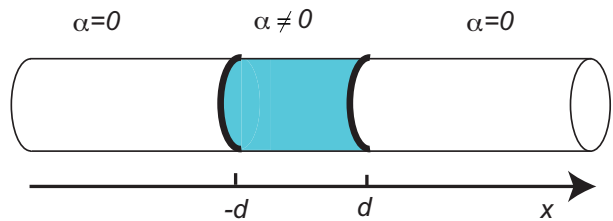}
\caption{(Color online) A schematic representation of our model: a hollow cylinder, containing a region, $|x|\leq d$, where a spin-orbit interaction, of strength $\alpha$, is active. The electrons are moving on the surface of the tube ballistically,  and thus
mode mixing is hindered. Short-range boundary scattering,  of  strength $\zeta$, separate the spin-orbit coupled region from the free regions.  The boundaries in-between these regions are marked by thick solid lines.
 }\label{SYST}
\end{figure}

In order to illustrate the general features of the scattering matrix as embodied in Eqs.  (\ref{NEW})-(\ref{VG}), we  study a specific example: transport through a two-dimensional stripe, of width $2d$ (along $x$), in which the electrons are subject to the Rashba spin-orbit interaction.
The
interfaces at $x=\pm d$ are  parallel to
the $y-$axis; effects of  interfacial scattering are  included
by   a repulsive delta-function potential located at
the interface. Elsewhere, the electrons move ballistically. Such a potential is characterized by a single
parameter, $\zeta$ (measured in momentum units). When $\zeta$ is very large, the interface
approaches the tunnel-junction limit. This model becomes particularly transparent when  a periodic boundary condition along the $y$ axis is assumed; then
the spin-orbit interaction does not mix the transversal modes. This model system, whose realization may be found in nanotubes (see in particular Ref. ~\onlinecite{SHACHAL}) is depicted in Fig. \ref{SYST}.
For cylinders, the coordinate $y$
represents the azimuthal angle around the cylinder (with perimeter ${\cal L}$), and the coordinate $z$
is radial, i.e. perpendicular to the surface of the cylinder.

The scattering matrix  is computed by solving for the scattering states of the Hamiltonian
\begin{align}
&{\cal H}=\frac{{\bf p}^{2}}{2m}+\frac{\zeta}{m}\Bigl (\delta (x+d)+\delta (x-d)\Bigr )\nonumber\\
&+\frac{\alpha}{2m}\Bigl (\Theta ( d-x) \Theta (x+d) {\bf p}\cdot
(\hat{z}\times\sig ) +{\rm hc}\Bigr )  \ ,
\end{align}
where  ${\bf p}$ is the (two-dimensional)
momentum operator and  $\alpha$ is the spin-orbit coupling (in momentum units).
With the periodic boundary condition, of period ${\cal L}$,  the wave functions vary along $y$ as $\exp[iqy]$, and the transverse momentum is quantized,
\begin{align}
q=\pm 2\pi n/{\cal L}\ ,\ \ \ n=0,\ 1,\ 2,\ \ldots\ . \label{Q}
\end{align}
Below, we measure all momenta (including $\alpha$ and $\zeta$) in units of $\hbar/{\cal L}$. The $x$ component of the wave vector is
\begin{align}
k=\sqrt{2mE-q^{2}}\ ,\ \ \ |x|\geq d\ ,\label{k}
\end{align}
for both spin components in the absence of the spin-orbit interaction, and
\begin{align}
k^{}_{\stackrel{u}{l}}&=\sqrt{(p^{}_{S}\pm\alpha )^{2}-q^{2}}\ ,
\label{kul}
\end{align}
for waves in the range $|x|<d$ where the spin-orbit interaction is effective.   \cite{silvestrov} Here,
\begin{align}
p_{S}=\sqrt{2mE+\alpha^{2}}\ .\label{PS}
\end{align}
As $q^{2}\le 2mE$, the wave vector $k_{u}$ is always {\em real};
in contrast, $k_{l}$ is purely imaginary for $|q|>p_{S}-\alpha$ (we assume $\alpha >0$),
and then one of the waves in the region $|x|<d$ is evanescent.
\cite{silvestrov,khodas}

Following the standard procedure of matching boundary conditions at the two interfaces $x=\pm d$ (allowing for the delta-function potentials \cite{sablikov,BC})
the scattering matrix of each channel takes the form
\begin{align}
{\cal S}=-1+\left [\begin{array}{cc}\cals^{}_{+}+\cals^{}_{-}&(\cals^{}_{+}-\cals^{}_{-})\sigma^{}_{x}\\
\sigma^{}_{x}(\cals^{}_{+}-\cals^{}_{-})&\ \ \ \sigma^{}_{x}(\cals^{}_{+}+\cals^{}_{-})\sigma^{}_{x}\end{array}\right ]\ ,\label{SDUG}
\end{align}
where
\begin{align}
\cals^{-1}_{\pm}=\frac{1}{k}
\Bigl (k+2i\zeta +iX^{}_{\pm}-i\sigma^{}_{z}(q-Z^{}_{\pm})+i\sigma^{}_{x}Y^{}_{\pm}\Bigr )\ . \label{sPM}
\end{align}
The functions $X_{\pm}$, $Y_{\pm}$, and $Z_{\pm}$ are all real,
independent of whether $k_{l}$ [see Eq. (\ref{kul})] is real or imaginary,
\begin{align}
&X_{\pm}^{}=\frac{p^{}_{S}}{D^{}_{0}}\Bigl (\frac{S^{}_{l} [C^{}_{u}(p^{}_{S}-\alpha )\pm q ]}{k_{l}^{}}+\frac{S^{}_{u} [C^{}_{l}(p^{}_{S}+\alpha )\mp q]}{k_{u}^{}}\Bigr )\ ,\nonumber\\
&Y^{}_{\pm}=\frac{p^{}_{S}}{D^{}_{0}}\Bigl (\frac{S^{}_{u} [qC^{}_{l}\mp (p^{}_{S}+\alpha ) ]}{k^{}_{u}}-\frac{S^{}_{l} [qC^{}_{u}\pm (p^{}_{S}-\alpha ) ]}{k^{}_{l}}\Bigr )\ ,\nonumber\\
&Z^{}_{\pm}=\frac{p^{}_{S}}{D^{}_{0}}\Bigl (2qp^{}_{S}\frac{S^{}_{l}S^{}_{u}}{k^{}_{l}k^{}_{u}}
\pm (C^{}_{l}-C^{}_{u})\Bigr )\ , \label{XYZ}
\end{align}
where
\begin{align}
D^{}_{0}=& 1-C^{}_{l}C^{}_{u}+\frac{p^{2}_{S}-\alpha^{2}_{}+q^{2}}{k^{}_{l}k^{}_{u}}S^{}_{l}S^{}_{u} \ . \label{det}
\end{align}
Here we use the shorthand notations
\begin{align}
S^{}_{l,u}\equiv\sin (2k^{}_{l,u}d )\ ,\ \ \ \ C^{}_{l,u}\equiv\cos (2k^{}_{l,u}d )\ .\label{SCLU}
\end{align}
When the transversal channels are not mixed, then the
total charge and spin transmissions are  given by
\begin{align}
{\cal T}=\sum_{n}{\cal T}_{nn}\ ,\ \ {\rm  and}\ \ \   {\bf V}=\sum_{n}{\bf V}_{nn}\ ,\label{TVNN}
\end{align}
see Eqs. (\ref{IR1MF}), (\ref{TT}), and (\ref{VV}). The scattering matrix Eq. (\ref{SDUG}) is unitary,  \cite{UNITARY}  and is self-dual \cite{BEEN} upon changing $q$ to $-q$.  (Other symmetries of the scattering matrix, in particular in the presence of the Zeeman interaction, are discussed in Refs. ~\onlinecite{ZHAISYMMETRIES} and ~\onlinecite{BRUSHEIM}.)
The self-duailty  symmetry of the scattering matrix implies that upon summing over all channels, as indicated by Eqs. (\ref{TVNN})  [i.e., over positive and negative values of the transverse wave vector $q$, see Eq. (\ref{Q})]  the contributions of the $x$ and $z$ directions of the polarization are cancelled, and  the net spin transmission  ${\bf V}$ is along the $y$ direction, normal to the direction of the current. \cite{silvestrov,ETO}

\begin{figure}[h ]
\includegraphics[width=8.5cm]{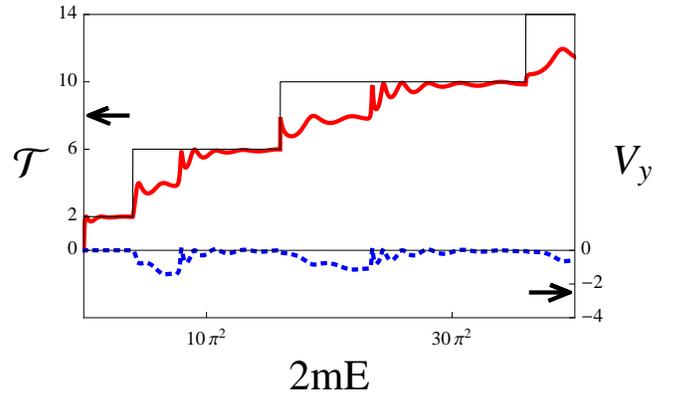}
\caption{(Color online) The transmission ${\cal T}$ through a
stripe with spin orbit interaction (thick line), the $y-$
component of the spin transmission $V^{}_{y}$  (dotted line), and
the staircase structure obtained in the absence of the spin-orbit
coupling (thin line) in a perfect system, as a function of the
energy. Here, the spin-orbit strength is $\alpha =0.9\pi$. }\label{CLEAN}
\end{figure}

\begin{figure}[h ]
\includegraphics[width=8.5cm]{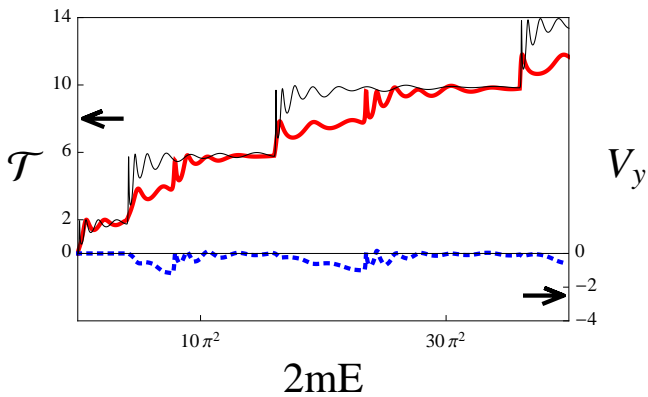}
\caption{(Color online) Same as Fig. \ref{CLEAN}, but in the presence of potential barriers at the boundaries, $\zeta =1.2$. Here, $\alpha =0.9\pi$.}\label{DIRTY}
\end{figure}

\begin{figure}[ h]
\vspace{1cm}
\includegraphics[width=8.5cm]{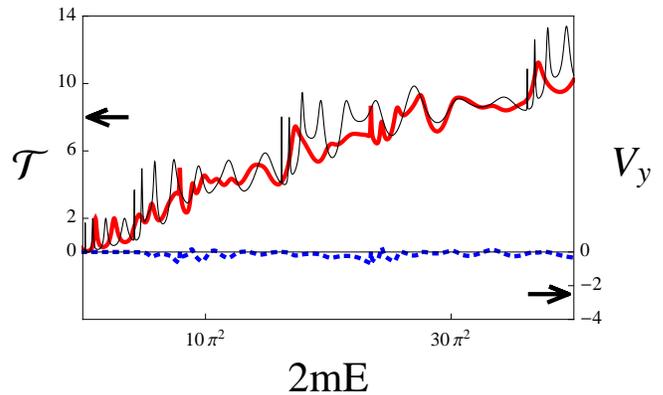}
\caption{(Color online) Same as Fig. \ref{CLEAN}, but with stronger boundary scattering,  $\zeta =4$, and $\alpha =.9 \pi$.}\label{NORA}
\end{figure}

Figures \ref{CLEAN}-\ref{NORA} were drawn using Eq. (\ref{SDUG}).  The curves in Fig. \ref{CLEAN} are computed for entirely  transparent boundaries ($\zeta =0$); indeed, the transmission in the absence of the spin-orbit coupling has the familiar perfect staircase structure, Eq. (\ref{BALLI}). Due to the choice of
periodic boundary conditions in the transverse direction, the
perfect transmission without the spin-orbit interaction becomes
$2N_{\perp}=2(2n+1)$, where $n=0,~1,~2,~\dots$ is the channel
number. The transmission computed in the presence of that coupling follows roughly this pattern,
but with two distinct new features.
Firstly, it shows interference oscillations. These arise since the spin-orbit interaction plays the role of a potential step at $|x|<d$. \cite{TINKHAM} One notes the complete vanishing of the spin conductance in  the leftmost part of the plot. There, the energy is too low to support a nonzero  transverse momentum $q$, and consequently the motion becomes effectively  one-dimensional. In such a case,     one may handle the effect of the spin-orbit interaction by a gauge transformation which multiplies the wave function by  $\exp[ 2ip_{S}d]$. Since this  phase factor is the sole effect of the spin-orbit interaction, the spin-polarization vanishes completely. \cite{kiselev} (Oscillations as a function of the spin-orbit coupling strength have been discussed in Ref. ~\onlinecite{CHAO}.)
When $q\neq 0$ and the motion restores its two-dimensional character,  this gauge transformation is no longer possible, and indeed  there appears spin conductance. The oscillations in the charge conductance,
resulting from the potential-step aspect of the spin-orbit interaction,  do persist.

The second prominent feature that the spin-orbit interaction
induces in the conductance pattern originates from the possibility to have evanescent waves.
At the beginning of each of the $n\neq 0$ steps [see Eq. (\ref{Q})], one of the waves pertaining to that energy becomes evanescent, since using Eq. (\ref{kul}) one has
\begin{align}
k^{}_{l}=i\kappa\
\end{align}
where
$\kappa $ is real. Then, $C_{l}$ and $-iS_{l}$ of Eqs. (\ref{SCLU})
are both of order $0.5\exp[2\kappa d]$, making the determinant of the transmission matrix ${\cal M}$, Eqs. (\ref{NEW}) and (\ref{SDUG}),  exponentially small. This implies
in turn that
one of the eigenvalues, Eqs. (\ref{MG}),   i.e. $\lambda^{-}_{nn}$, is very small. As a result, it follows from Eq. (\ref{TT}) that
the (perfect)
transmission is reduced to ${\cal
T}_{nn}\simeq\lambda^{+}_{nn}\simeq 1$. Concomitantly, the
polarization magnitude, $|{\bf V}_{nn}(q)|$, [see Eq.  (\ref{VG})] is also equal to
$\lambda^{+}_{nn}\simeq 1$.  However,  upon adding  ${\bf
V}(q)+{\bf V}(-q)$ the resulting vector ${\bf V}$ is along the
$y-$axis, as shown in Fig. \ref{CLEAN}. This
$y-$component decreases gradually from 0 towards $-1$ as the
energy increases in the steps where ${\cal T}\simeq 4n$. At higher
energies, both $k_{u}$ and $k_{l}$ are real, and the (perfect)
transmission becomes ${\cal T}_{nn}=\lambda^{+}_{nn}+\lambda
^{-}_{nn}\approx 2$, while the spin transmission becomes  $|{\bf
V}_{nn}(q)|=\lambda^{+}_{nn}-\lambda^{-}_{nn}\approx 0$.  For the
parameters used in  Figs. \ref{CLEAN} and \ref{DIRTY}, the
contributions of all the steps except the highest one are in this
latter regime, and therefore each of them contributes 2 (if $q=0$) or 4 (if $q\ne 0$)  to
the transmission and 0 to the spin transmission. Thus, the charge
conductance  exhibits $2(2n+1)$ and $4n$  multiples of $e^2/h$,
while the spin transmission is roughly  between 0 and $-1$. For
higher values of $\alpha$, one of the waves may remain evanescent
until the next step begins, yielding only  $4n$  steps in the
transmission. \cite{silvestrov} In our dimensionless units, this condition becomes $\alpha >(2n+1)\pi /n$.

Repeating the computation for finite barriers at the two
interfaces [see Fig. \ref{DIRTY}] shows that  a modest amount of
interface scattering  is not detrimental, and the two main
features discussed above are still detected. Even more interesting
are the curves shown in Fig. \ref{NORA}. Here, the amount of
scattering at the interfaces has been increased such that  the
staircase structure of the charge conductance, both in the
presence and in the absence of the spin-orbit coupling, is
smeared;  since the conductance is not quantized, the
spin-polarization is small (but still negative) for all values of
the energy.

\vspace{1cm}

\section{Concluding remarks}
\label{DISCU}

In conclusion,  we have derived a Landauer formula for the charge and the spin conductances.  The correlations
between the two conductances and their quantization have been demonstrated for transport through cylinders with stripes which have nonzero
spin-orbit interactions. Measurements of the  charge transmission can thus yield information on the spin
polarization. It would be interesting to check these predictions experimentally, in particular on the new set-ups   of core-shell nanowires or  made of carbon nanotubes.

\begin{acknowledgments}

We thank P. Silvestrov for helpful comments. AA and OEW acknowledge support by the binational Israel-United States foundation,   by a grant from  the German Federal
Ministry of Eduction and Research (BMBF) within the framework of
the German-Israeli Project Cooperation (DIP), and the
hospitality of NTT Basic Research Laboratory at Atsugi-shi,
Japan. YA acknowledges partial support from the ISF, and  the hospitality of the Graduate School of
Mathematical Science at the University of Tokyo.

\end{acknowledgments}


\begin{thebibliography}{999}

\bibitem{landauer}

R. Landauer, IBM J. Res. Dev. {\bf 1}, 233 (1957); Philos. Mag. {\bf 21}, 863 (1970).

\bibitem{LANDAUER}
Y. Imry and R. Landauer, Rev. Mod. Phys. {\bf 71}, S306 (1999).

\bibitem{book} Y. Imry, {\it Introduction to Mesoscopic Physics}, 2nd ed. (Oxford
University Press, Oxford, 2002).

\bibitem{wharam}
D. A. Wharam, T. J. Thornton, R. Newbury, M. Pepper, H. Ahmed, J.
E. F. Frost, D. G. Hasko, D. C. Peacock, D. A. Ritchie, and G. A.
C. Jones, J. Phys. {\bf C21}, L209 (1988);
B. J. van Wees, H. van Houten, C. W. J. Beenakker, J. G.
Williamsom, L. P. Kouwenhoven, D.  van der Marel, and C. T. Foxon,
Phys. Rev. Lett. {\bf 60}, 848 (1988).


\bibitem{buttiker}


M. B\"{u}ttiker, Phys. Rev. B {\bf 46}, 12485 (1992).







\bibitem{levinson} Y. Levinson, Phys. Rev. B {\bf 61}, 4748
(2000).



\bibitem{niu}

J. Shi, P. Zhang, D. Xio, and Q. Niu, Phys. Rev. Lett. {\bf 96}, 076604 (2006).


\bibitem{NUD}

B.  Nikoli\'{c}, L. P. Z$\hat{a}$rbo,  and S. Souma, Phys. Rev. B {\bf 73}, 075303 (2006).
\bibitem{MOROZ}
A. V. Moroz and C. H. W. Barnes, Phys. Rev. B {\bf 60}, 14272 (1999).

\bibitem {PRIVMAN}

Y. V. Pershin, J. A. Nesteroff, and V. Privman, Phys. Rev. B {\bf 69}, 121306(R) (2004).

\bibitem{silvestrov}
P. G. Silvestrov and E. G. Mischenko, Phys. Rev. B {\bf 74},
165301 (2006).
\bibitem{rashba}

E. I. Rashba,  Fiz. Tverd. Tela (Leningrad) {\bf 2}, 1224 (1960)
[Sov. Phys. Solid State {\bf 2}, 1109 (1960)]; Yu. A. Bychkov and
E. I. Rashba, Pis'ma Zh. Eksp. Teor. Fiz. {\bf 39}, 66 (1984)
[Sov. Phys. JETP Lett. {\bf 39}, 78 (1984)].


\bibitem{DRES}
G. Dresselhaus, Phys. Rev. {\bf 100}, 580 (1955).

\bibitem{Bert}
J. J. Krich and B. I. Halperin, Phys. Rev. Lett. {\bf 98}, 226802
(2007).





\bibitem{CUBIC}

A. A. Kovalev, M. F. Borunda, T. Jungwirth, L. W. Molenkamp, and
J. Sinova, Phys. Rev. B {\bf 76}, 125307 (2007).

\bibitem{nitta}

T. Koga, J. Nitta, T. Akazaki, and H. Takayanagi, Phys. Rev. Lett.
{\bf 89}, 046801 (2002).


\bibitem{SO1}

M. K\"{o}nig, A. Tschetschekin, E. M. Hankiewicz, J. Sinova, V.
Hock, V. Daumer, M. Sch\"{a}fer, C. R. Becker, H. Buhmann, and L.
W. Molenkamp, Phys. Rev. Lett. {\bf 96}, 076804 (2006).


\bibitem{SO2}
T. Bergsten, T. Kobayashi, Y. Sekine, and J. Nitta, Phys. Rev.
Lett. {\bf 97}, 196803 (2006).

\bibitem{sheka}

E. I. Rashba and V. I. Sheka, in {\it Landau Level Spectroscopy}, G. Landwehr and E. I. Rashba, eds.
(Elsevier, Amsterdam, 1991).

\bibitem{refereeb}
B. Nikoli\'{c} and S. Souma, Phys. Rev. B {\bf 71}, 195328 (2005);
R. L. Dragomirova and B. K. Nikoli\'{c}, Phys. Rev. B {\bf 75}, 085328 (2007).

\bibitem{sablikov}

V. A. Sablikov, A.   A. Sukhanov,   and Y. Ya. Tkach, Phys. Rev. B {\bf 78}, 153302 (2008).

\bibitem{khodas}

M. Khodas, A. Shekhter, and A. M. Finkel'stein, Phys. Rev. Lett.
{\bf 92}, 086602 (2004); A. Shekhter, M. Khodas, and A. M.
Finkel'stein, Phys. Rev. B {\bf 71}, 125114 (2005).


\bibitem{BERCIOUX}


V. Marigliano Ramaglia, D. Bercioux, V. Cataudella, G. De Filippis, and C. A. Perroni, J. Phys.: Condens. Matter {\bf 16}, 9143 (2004).


\bibitem{halperin1}

I. Adagideli, G. E. W. Bauer, and B. I. Halperin, Phys. Rev. Lett. {\bf 97}, 256601 (2006).





\bibitem{MIRELES}


F. Mireles and G. Kirczenow, Phys. Rev. B {\bf 64},  024426 (2001).

\bibitem{ZHU}

Shi-Liang Zhu,
Z. D. Wang, and
Lian Hu,
J. Appl. Phys. {\bf 91}, 6545 (2002).



\bibitem{Governale}


M. Governale and U. Z\"{u}licke, Solid State Comm. {\bf 131}, 581 (2004).





\bibitem{ETO}

M. Eto, T. Hayashi, and Y. Kurotani, J. Phys. Soc. Jpn. {\bf 74}, 1934 (2005).

\bibitem{ANDO}

The spin-orbit interaction in nanotubes is not predominantly of the Rashba type used in Sec. \ref{CYLINDER}, see T. Ando, J. Phys. Soc. Jpn. {\bf 69}, 1757 (2000). However, the results of Sec. \ref{LAN} 
are valid for these systems.




\bibitem{SHACHAL}


F. Kuemmeth, S. Ilani, D. C. Ralph, and P. L. McEuen, Nature {\bf 452}, 448 (2008).


\bibitem{CORE}


L. J. Lauhon, M. S. Gudiksen, D. Wang, and C.
M. Lieber,  Nature {\bf 420}, 57 (2002);
J. Noborisaka, J. Motohisa, S. Hara, and T. Fukui,  Appl. Phys. Lett. {\bf 87},
093109 (2005);
G. Zhang, K. Tateno, T.  Sogawa, and H.
Nakano, Appl. Phys. Express {\bf 1}, 063003 (2008);
J. W. W. van Tilburg, R. E. Algra, W. G. G. Immink, M. Verheijen,
E. P. A. M. Bakkers, and L. P. Kouwenhoven, Semicond. Sci. Technol. {\bf 25}, 024011 (2010).



\bibitem{COMMENT1}

The extension of the formalism presented in Sec. \ref{LAN} to the case of spin-polarized leads is rather straightforward.




\bibitem{BC}
When writing down the boundary conditions,
one should  match the covariant momentum--which includes the spin-orbit contribution.

\bibitem{UNITARY}
Since $\cals^{-1}_{\pm}+(\cals^{\dagger}_{\pm})^{-1}=2$,
the scattering matrix Eq. (\ref{SDUG}) is unitary.


\bibitem{BEEN}

C. W. J. Beenakker, Rev. Mod. Phys. {\bf 69}, 731 (1997).


\bibitem{ZHAISYMMETRIES}

F. Zhai and H. Q. Xu, Phys. Rev. Lett. {\bf 94}, 246601 (2005).




\bibitem{BRUSHEIM}

P. Brusheim and H. Q. Xu, Phys. Rev. B {\bf 74}, 233306 (2006).

\bibitem{TINKHAM}

This effect resembles in a way the appearance of  Andreev reflections at the interface between a superconductor and a normal metal, see G. E. Blonder, M. Tinkham, and T. M. Klapwijk, Phys. Rev. B {\bf
25}, 4515 (1982).


\bibitem{kiselev}



A. A. Kiselev and K. W.  Kim,  Phys. Rev. B {\bf 71}, 153315 (2005)]; note that  S\'{a}nchez and Serra [Phys. Rev. B {\bf 74}, 153313 (2006)] do not find polarization for an harmonic confining potential along $y$.


\bibitem{CHAO}


A. G. Mal'shukov, V. V. Shlyapin, and K. A. Chao, Phys. Rev. B {\bf 66}, 081311 (2002); C-H Chang, A. G. Mal'shukov, and K. A. Chao, Phys. Lett. A {\bf 326}, 436 (2004).

















\end{thebibliography}
\end{document}